\documentclass[a4paper,11pt]{article}
\usepackage{pos}
\usepackage[]{caption}
\usepackage{subcaption}
\usepackage{amsmath,amssymb}

\title{Analysis framework for Multi-messenger Astronomy with IceCube}

\ShortTitle{Analysis framework for Multi-messenger Astronomy with IceCube}

\author{The IceCube Collaboration \\{\normalsize \normalfont(a complete list of authors can be found at the end of the proceedings)}}
\emailAdd{klfan@umd.edu}

\abstract{Combining observational data from multiple instruments for multi-messenger astronomy can be challenging due to the complexity of the instrument response functions and likelihood calculation. We introduce a python-based unbinned-likelihood analysis package called i3mla (IceCube Maximum Likelihood Analysis). i3mla is designed to be compatible with the Multi-Mission Maximum Likelihood (3ML) framework, which enables multi-messenger astronomy analyses by combining the likelihood across different instruments. By making it possible to use IceCube data in the 3ML framework, we aim to facilitate the use of neutrino data in multi-messenger astronomy. In this work we illustrate how to use the i3mla package with 3ML and present preliminary sensitivities using the i3mla package and 3ML through a joint-fit with HAWC Public dataset.

\vspace{4mm}
{\bfseries Corresponding authors:}
Kwok Lung Fan$^{1*}$, John Evans$^{1}$, Michael Larson$^{1}$\\
{$^{1}$ \itshape University of Maryland}\\[4mm]
$^*$ Presenter
\FullConference{37$^{\rm{th}}$ International Cosmic Ray Conference (ICRC 2021)\\
		July 12th -- 23rd, 2021\\
		Online -- Berlin, Germany}

}
\begin{document}
\maketitle

\section{Introduction}\label{sec:info}
IceCube is a cubic-kilometer scale Cherenkov neutrino detector operating at the South Pole\cite{Aartsen:2016nxy}. It consists of over 5000 DOMs (digital optical modules), each containing a 10-inch photomultiplier tubes(PMT). Completed in 2010, IceCube has been continuously observing astrophysical neutrinos for more than ten years, opening a new window into the universe. However, combining the IceCube neutrino data with other observational data from multiple instruments for multi-messenger has always been non-trivial, due to the differences in the likelihood calculation. Here, we introduce a new python-based unbinned likelihood analysis package called IceCube Maximum Likelihood Analysis (i3mla). i3mla aims to solve this problem by providing a framework to make the IceCube public data compatible with the Multi-Mission Maximum Likelihood (3ML) architecture \cite{Vianello:2015tuw}.\\

\section{Likelihood Formalism}\label{sec:likelihood}
\label{sec:likelihood}
The analysis technique for IceCube data used in i3mla is the standard unbinned likelihood method that has been widely used in both IceCube and other experiments. The construction of the likelihood in i3mla is similar to that of the IceCube TXS 0506+056  analysis \cite{aartsen2018neutrino}. The likelihood value of each neutrino event is defined as:
\begin{equation}    
\begin{split}
L_i(\vec{\theta},\vec{D_i}) = \frac{n_s}{N}S(\vec{\theta},\vec{D_i}) + \frac{N-n_s}{N}B(\vec{\theta},\vec{D_i}).
\end{split}
\end{equation}
Hence the total likelihood is the multiplication of the likelihood value of the individual neutrino event.Here, N is the total number of neutrino events, $\vec{D_i}$ represents the neutrino event properties (direction, reconstructed energy, etc.), $\vec{\theta}$ is the parameters of the physical model, including source location, morphology, energy spectrum, etc. and $n_s$ is the number of signal events that maximizes the likelihood. $S$ and $B$ are the signal probability density function (pdf) and the background probability density function which are a function of the neutrino properties and the source model.\\
\begin{equation}    
\begin{split}
L = \prod_i^N L_i(\vec{\theta},\vec{D_i}) = \prod_i^N\frac{n_s}{N}S(\vec{\theta},\vec{D_i}) + \frac{N-n_s}{N}B(\vec{\theta},\vec{D_i}).
\end{split}
\end{equation}

In the absence of any neutrino signal, we can define the null model by setting the $n_s$ to 0. Then we construct construct the log-likelihood ratio between the signal hypothesis and the background-only hypothesis and use it as a test statistic for the likelihood ratio test. It is convenient to multiply the log-likelihood ratio by 2 such that Wilk's theorem applies and the resulting test statistic value will approximately follow a $\chi^2$ distribution with a degree of freedom equal to the number of free parameters \cite{wilks1938large}. However, it is worth noticing that the $\chi^2$ distribution is only an approximation to our TS distribution. All IceCube analyses estimate the significance and p-value by simulating background-only trials ($n_s=0$) and building the background TS distribution. The final test statistic is of the form:
\begin{equation}    
\begin{split}
TS &= 2\sum_i^N log \left(\frac{L_i(\vec{\theta},\vec{D_i})}{L_i(n_s = 0,\vec{D_i}))}\right)\\ 
&= 2\sum_i^N log\left( \frac{n_s}{N}(\frac{S(\vec{D_i},\vec{\theta})}{B(\vec{D_i},\vec{\theta})}-1)+1\right).
\end{split}
\end{equation}
The signal and background pdf consist of multiple different pdf terms and can be modified for different analyses. The commonly used signal pdf and background pdf for IceCube analyses consist of spatial, energy, and temporal terms \cite{aartsen2018neutrino}. The signal spatial pdf under the point source hypothesis is modeled as a 2D Gaussian with:
\begin{equation}    
\begin{split}
S_{spatial}(\vec{r},\vec{r}_{\nu},\sigma)&=\frac{1}{2\pi \sigma^2}e^{-\frac{(\vec{r}-\vec{r}_{\nu})^2}{2\sigma^2}}.
\end{split}
\end{equation}
where $\vec{r},\;\vec{r}_{\nu},\;\sigma$ is the location of the source, the reconstructed direction of the event and the estimated angular error of the event. For extended source:
\begin{equation}    
\begin{split}
S_{spatial}(f(\vec{r}),\vec{r}_{\nu},\sigma)&=\int_{\mathbb{S}^2} f(\vec{r})\frac{1}{2\pi \sigma^2}e^{-\frac{(\vec{r}-\vec{r}_{\nu})^2}{2\sigma^2}}d\vec{r} \\
&\approx \int_{x}\int_{y}  f(\vec{r})\frac{1}{2\pi \sigma^2}e^{-\frac{(\vec{r}-\vec{r}_{\nu})^2}{2\sigma^2}}dxdy,
\end{split}
\end{equation}
where $f(\vec{r})$ is the spatial pdf of the extended source. The integral can be approximated numerically by dividing the sky into a grid and summing the contribution of each grid cell:
\begin{equation}    
\begin{split}
S_{spatial}(g(\vec{r}),\vec{r}_{events},\sigma)
&\approx \sum_{i}  g(\vec{r}_i)\frac{1}{2\pi \sigma^2}e^{-\frac{(\vec{r_i}-\vec{r}_{\nu})^2}{2\sigma^2}},
\end{split}
\end{equation}
where $g(\vec{r}_i)$ is the spatial probability mass function of the extended source and $\vec{r}_i$ is the pixel center. For a Gaussian extended source with extension $\sigma_A$, the above formula can be simplifed into:
\begin{equation}    
\begin{split}
S_{spatial}(\vec{r}_{src},\vec{r}_{events},\sigma,\sigma_A)=\frac{1}{2\pi (\sigma^2+\sigma_A^2)}e^{-\frac{(\vec{r}_{src}-\vec{r}_{events})^2}{2(\sigma^2+\sigma_A^2)}}.
\end{split}
\end{equation}
The background pdf is modeled as:
\begin{equation}    
\begin{split}
B_{spatial}(\vec{r}_{events})=\frac{1}{2\pi }\delta(dec),
\end{split}
\end{equation}
which only depends on the reconstructed declination since the component along right ascension is assumed to be uniform due to detector geometry. The $\delta(dec)$ distribution can be constructed from background data.\\ 

The signal energy pdf is computed using Monte Carlo simulation to construct a source spectrum hypothesis through a 2D histogram binned in reconstructed declination and reconstructed energy bin. For each bin, we compute the expected number of signal events given the source model and then normalize over the reconstructed declination bin. An alternative way to construct the signal energy pdf is to use the effective area at a certain true energy and reconstructed declination. Multiplying the effective area with the differential flux at that true energy  then returns the expected number of neutrinos at that true energy. Then the energy smearing matrix can be applied to get the distribution of reconstructed energy. Repeating the process for all bins and we can obtain the final histogram. This can be used when only tabulated instrument response functions (IRFs) are available. For the background energy pdf, we simply use background data to construct the 2D histogram and normalize over reconstructed declination. For the highest energy bins which have low statistic, we use the pdf value at the nearest reconstructed energy bin to avoid an infinity signal-over-background value.\\

The signal time pdf term is the source emission model over time. For continuous-emission or box-shape model, the pdf is a uniform distribution $S(t) = \frac{1}{T_w}$ where $T_w$ is the time window. The background time pdf is usually also modeled as a uniform distribution \cite{aartsen2018neutrino}. An overview of the code structure is shown in Figure 1.\\

\section{i3mla framework}\label{sec:framework}
i3mla is a likelihood framework for IceCube. It is implemented entirely with python and uses the new python 3.8 features for modularized design, dataclasses management, and shared memory parallel-processing. i3mla itself can work as a standalone package for IceCube analyses with fast trial generation and fast TS minimization features. It is also fully compatible with the Multi-Mission Maximum Likelihood framework (3ML) through a custom IceCube plugin and provides a simple interface for multi-messenger analyses.\\
\begin{figure}[t]
\includegraphics[width=\textwidth]{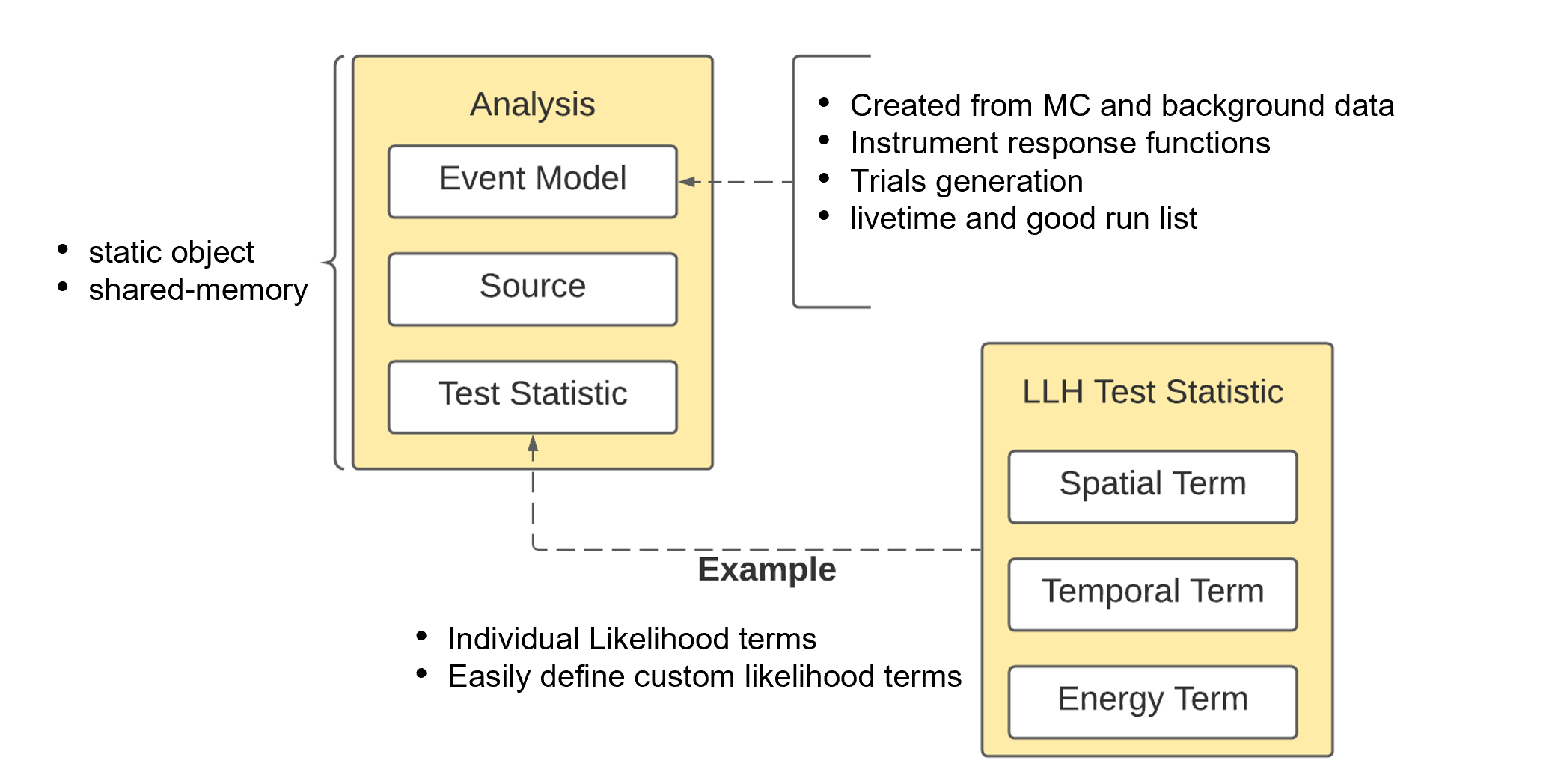}
\caption{Code structure of i3mla}
\end{figure}
The `Analysis' class is the core class of i3mla and handles the likelihood calculation and the trial generation. It consists of three different classes: `Source', `Event Model', and `Test Statistic'. The `Analysis' object will be created in the backend when using 3ML interface.
\subsection{Source model}
i3mla currently supports both point source source models which contains the location (right accession and declination) of the source and extended source models. Gaussian extended source models can be defined with parameter $\sigma_A$. A general extended source can be defined with skymap grid with the pixels values representing the weights which sum to 1.

The injection spectrum and spectral hypothesis of the point source are unrestricted and can be defined through the astromodels package when using the 3ML interface. However, it will slow the minimization process as the signal energy pdf is calculated during the fit. When using i3mla as a standalone package, only a power law over IceCube's energy range can be used as a spectral hypothesis, but it will greatly increase the minimization speed as the signal energy pdf is pre-calculated.
\subsection{Test Statistic}
The test statistic object consists of the different signal and background terms described in section 2. For the most common IceCube analyses, it consists of spatial, energy and temporal terms. The code is designed to be highly modularized. Users can choose to include or exclude individual likelihood terms and define a new likelihood term customized for a specific analysis. 
\subsection{Event Model}
Event Model is an analogy of Instrument Response Functions (IRFs) for IceCube. Creating an Event Model object requires Monte Carlo simulation, real data as background, and also the good run list which stores the livetime and the total number of events in each run (usually 8 hours of detector uptime). The Event Model calculates the expected number of neutrinos $n_s$ in Eq. (2.3) and the signal energy pdf given the flux model using the Monte Carlo and calculates the background spatial pdf in Eq. (2.5) using the data as background. \\
Besides calculating the likelihood value, Event Model also handles signal injection and trial generation. Given the injection flux model and source location, Event Model first calculates the expected number of neutrinos using Monte Carlo, then sets the injected number of neutrinos by drawing a number from a Poisson distribution with the mean equal to the expected number of neutrinos. Monte Carlo events are then drawn from the declination band of interest based on their true energy, effective area, and flux. Events are then rotated such that the true right ascension and declination coincide with the source location. For background, we draw background events based on the event rate and shuffle the right accession and arrival time of the events.

\section{Validation and Testing}

\subsection{Signal bias and spectral bias}
Being able to fit the true flux normalisation and true spectral information is important for multi-messenger astronomy, especially in constraining physics models. Here we inject a power law $E^{\gamma}$ spectrum with a range of spectral indices at the location of TXS 0506+056 using a 3 year of time window and fit it with a power law at the exact location to test for a potential bias in the normalisation and spectral index of the best-fit parameters.\\
\begin{figure}[h!]
\centering
\begin{subfigure}{.5\textwidth}
  \centering
  \includegraphics[width=\linewidth]{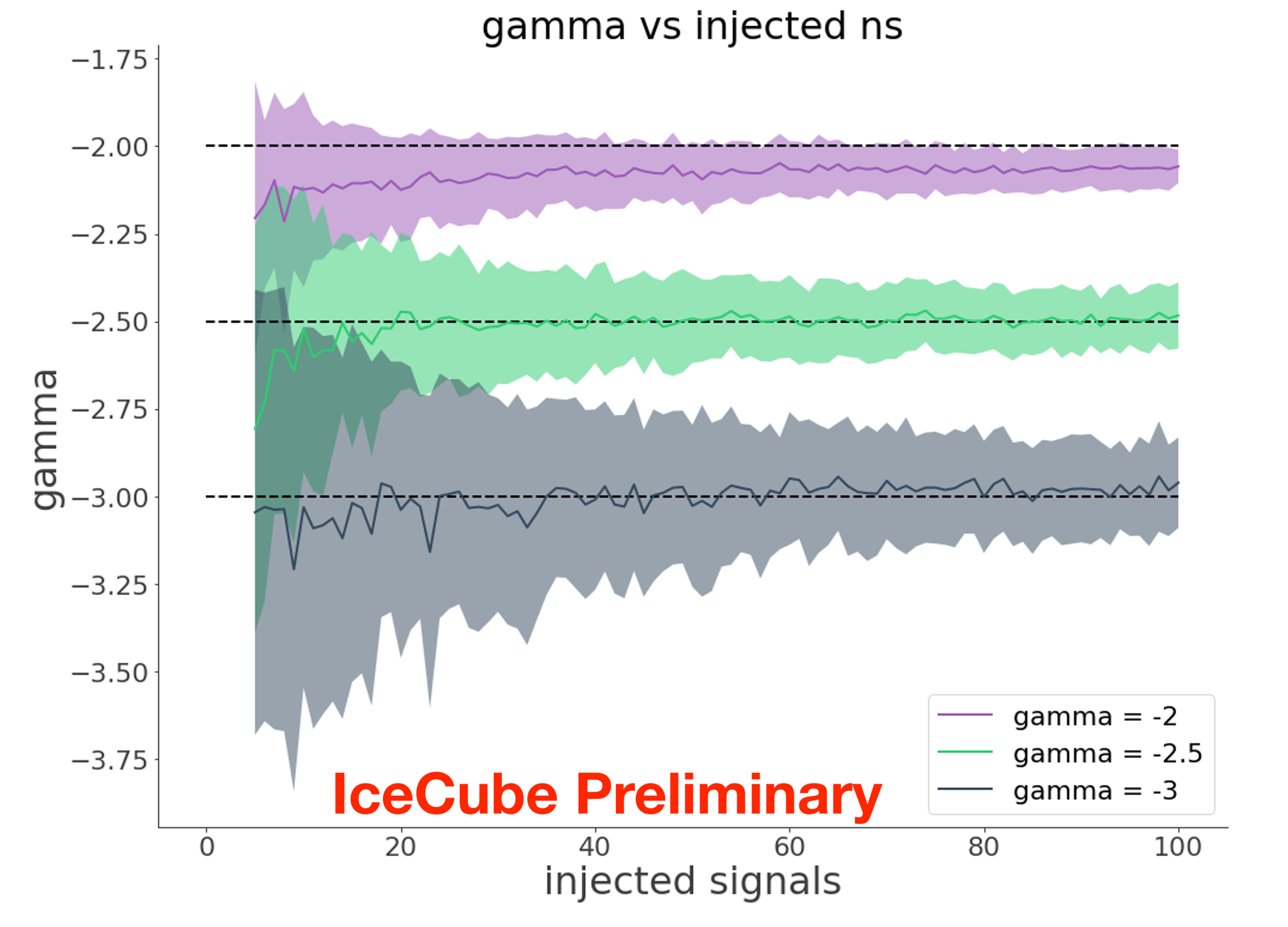}
  \caption{index bias plot}
  \label{fig:sub1}
\end{subfigure}%
\begin{subfigure}{.5\textwidth}
  \centering
  \includegraphics[width=\linewidth]{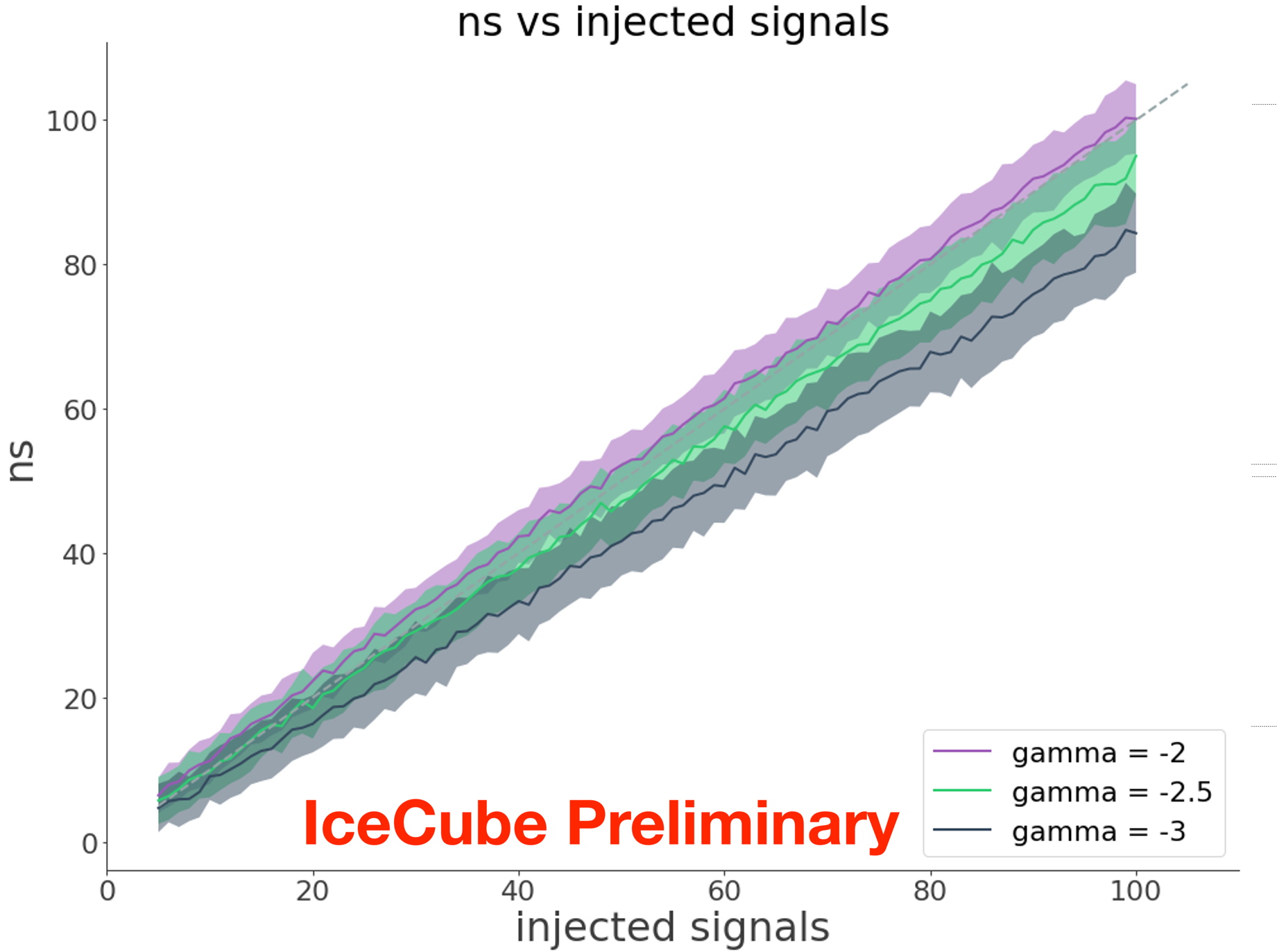}
  \caption{ns bias plot}
  \label{fig:sub2}
\end{subfigure}
\caption{Best fit ns and spectral index for different spectral indices as a function of injected signal. }

\label{fig:test}
\end{figure}

Figure 2.a shows the mean best-fit index and 1 sigma uncertainty region (100 tests) for the different spectral indices as a function of the number of injected neutrinos events. The uncertainty goes down and the mean best fit index coincides with injected index when the signal is strong. The spectral index of -2 shows a bias toward a softer index even when the signal is strong. Figure 2.b shows the mean best fit $n_s$ and 1 sigma uncertainty region for the different indices as a function of the injected number of neutrinos. For a spectral index of -2, the best fit $n_s$ coincides with the true number of injected neutrinos. The spectral index of -2 and -2.5 tend to underestimate the number of injected neutrinos.\\

This phenomenon also exists in other IceCube analysis software and the root of this bias is believed to be the mis-modeling of the spatial signal pdf. The point spread function (PSF) of IceCube is traditionally modeled as a 2D Gaussian but Monte Carlo simulation shows that the PSF usually has a long tail and hence deviates from the Gaussian approximation.  Methods of reducing this effect are currently being studied by IceCube Collaboration and will be applied to i3mla in future development.

\subsection{Identification of flare}
Blazar TXS 0506+056 is the first $>3\sigma$ source identified by IceCube \cite{aartsen2018neutrino}. We inject a source at the TXS 0506+056 location with a spectral index of -2.1, time integrated flux $E^2F = 2.1 \times 10^{-4} \; \mathrm{ TeVcm^{-2}}$ at 100 TeV and a Gaussian time window centered at MJD 57004 with $\sigma = 110$ days. These parameters are the same as the IceCube best-fit result of the TXS 0506+056 flare \cite{aartsen2018neutrino}. We use a box-shaped signal time profile and select candidate time windows by all the possible combinations of events with a spatial Signal/Background likelihood higher than 1000. From these the time windows with the highest TS-value is then selected. Considering the computational resource required, we use i3mla as a standalone package to run the flare search which is 200 times faster than using 3ML.
\begin{figure}[h!]
\centering
\begin{subfigure}{.5\textwidth}
  \centering
  \includegraphics[width=\linewidth]{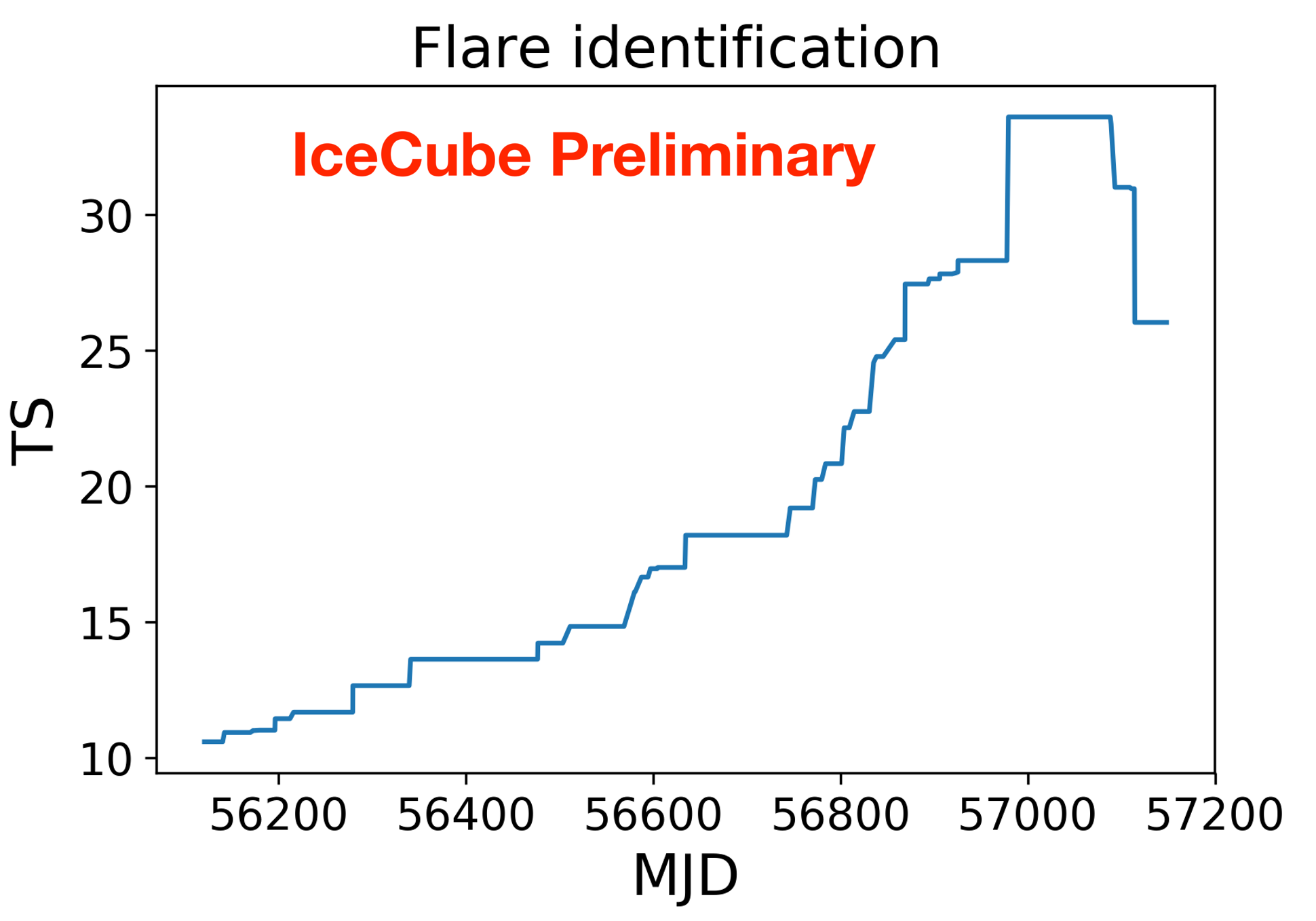}
  \caption{Flare identification of one example trial}
  \label{fig:sub1}
\end{subfigure}%
\begin{subfigure}{.5\textwidth}
  \centering
  \includegraphics[width=\linewidth]{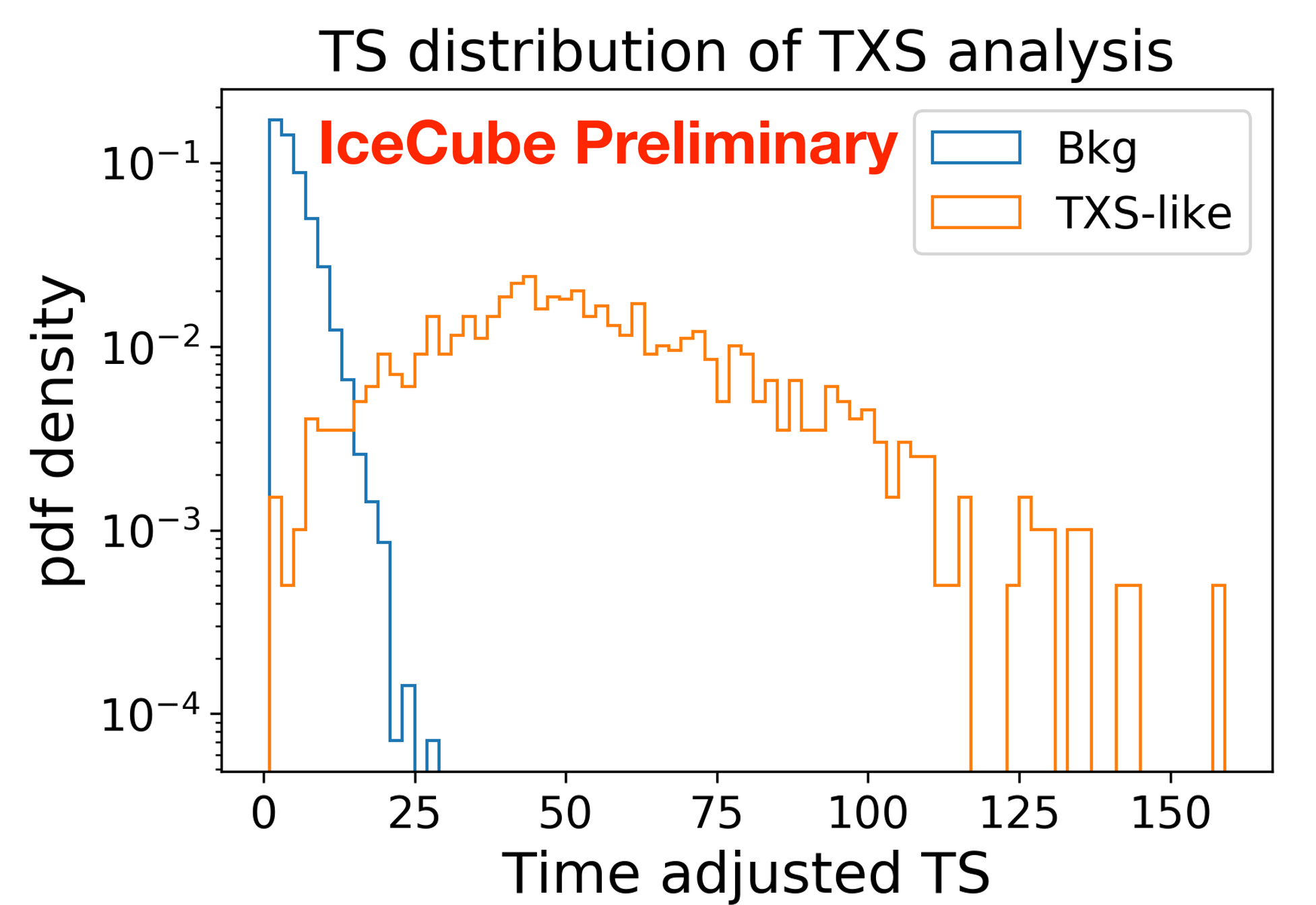}
  \caption{TS distribution }
  \label{fig:sub2}
\end{subfigure}
\caption{The left plot shows the flare identification of one example trial. The $x$-axis is time and $y$-axis is the best TS from the time window which contains that time, hence the plateau in the plot with highest TS is the flare period identified.The right plot show TS distribution of background (blue) and with TXS 0506+056 flare-like spectrum injected (orange). }

\label{fig:test2}
\end{figure}
Figure 3a shows the success of identifying the flaring period for an example trial. Figure 3b shows the background TS distribution and the TS distribution with signal injected. The flare can be discovered with $3 \sigma$ with probability over $80\%$ of the time.

\subsection{Joint-fit with HAWC Crab public dataset}
The High Altitude Water Cherenkov Observatory (HAWC) is a ground-based water Cherenkov observatory completed in 2015. HAWC observes very-high-energy gamma rays up to 100 TeV with a long duty cycle and wide field-of-view. The overlap in energy range between IceCube Neutrino events and HAWC gamma-ray events enables multi-messenger search for common origin between neutrinos and gamma-rays. We use HAWC public Crab data which was taken from 2014 November 26 to 2016 June 2, a total detector livetime of 507 days \cite{abeysekara2017observation}. We inject a neutrino source at the Crab with the same time period. The livetime of IceCube at that period is 545 days. The injection spectrum has the same shape as HAWC's latest Crab spectrum, a log-parabola 
\begin{equation}    
\begin{split}
\frac{dN}{dE}=\phi_0 \left(\frac{E}{7~\mathrm{TeV}}\right)^{(-\alpha -\beta \ln{(E/7~\mathrm{TeV})}}
\end{split}
\end{equation}
with $\alpha = 2.79$ and $\beta = 0.1$ \cite{abeysekara2019measurement}. We vary the flux norm level, from 5\% of the Crab Nebula flux to 100\% of the Crab flux ($\phi_0 = 2.13 \times 10^{-13} \; \mathrm{TeV cm^{-2} s^{-1}}$) to find the sensitivity\footnote{90\% signal trials TS > median of background TS distribution.} and the discovery potential\footnote{Median of the signal trials TS> $3 \sigma$ of background TS distribution.}. We perform a joint-fit using 3ML with a log-parabola. We fixed the spectral shape of the neutrinos and gamma-rays ($\alpha$ and $\beta$) to be the same and allow the flux normalization to float.
\begin{figure}[h!]
\centering
\begin{subfigure}{.5\textwidth}
  \centering
  \includegraphics[width=\linewidth]{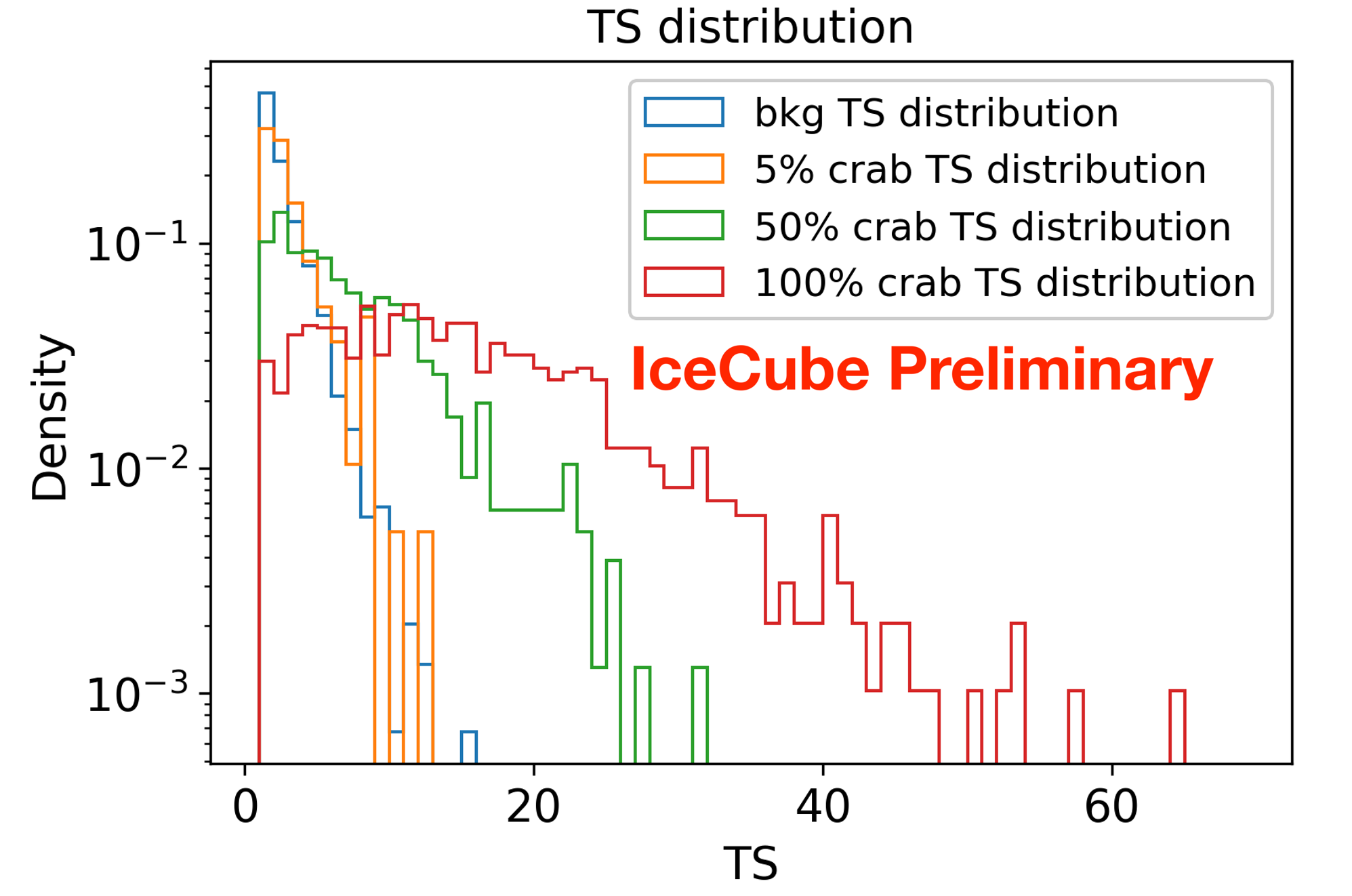}
  \caption{TS distribution of different level of crab flux}
  \label{fig:sub1}
\end{subfigure}%
\begin{subfigure}{.5\textwidth}
  \centering
  \includegraphics[width=\linewidth]{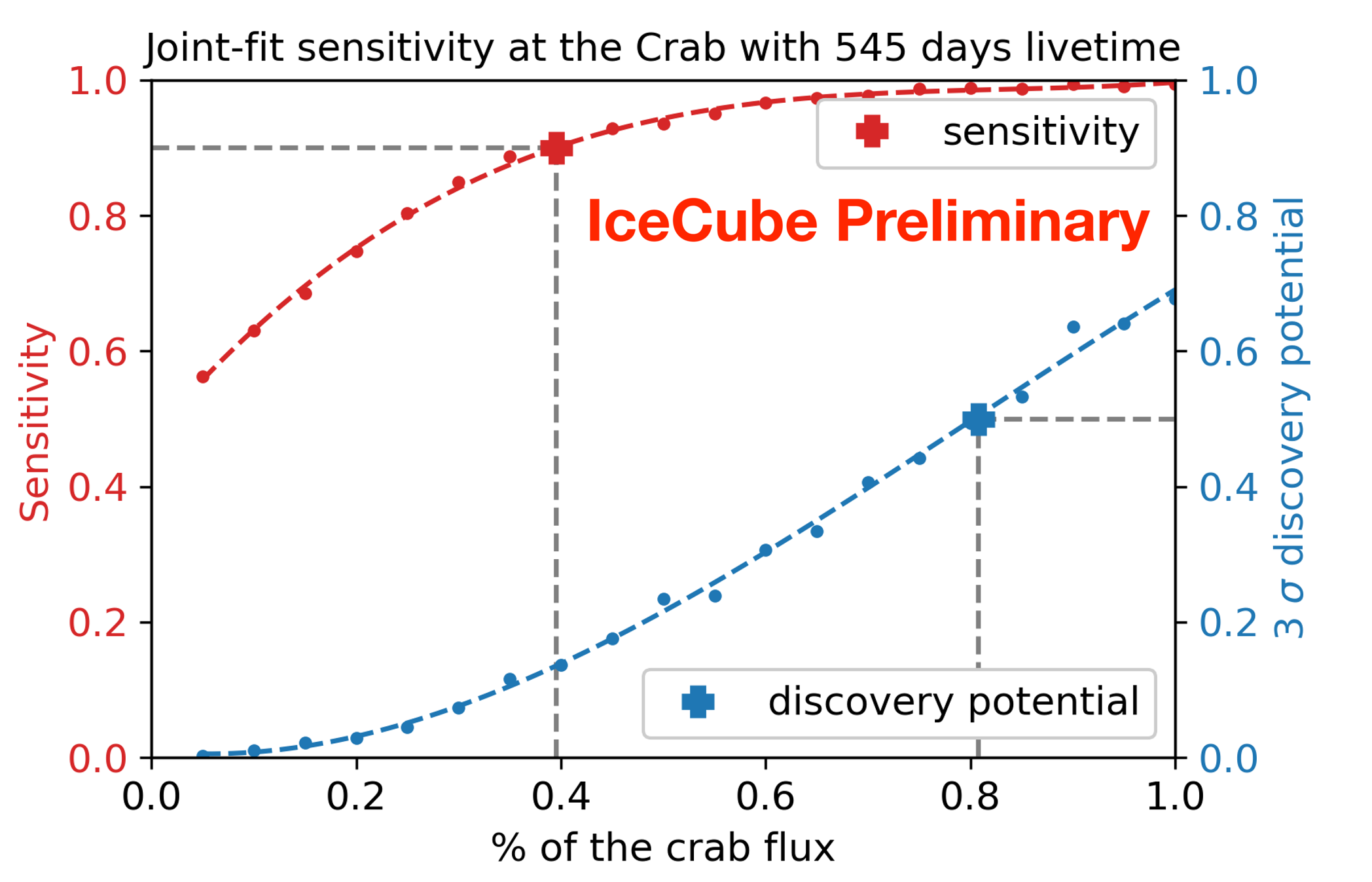}
  \caption{Sensitivity and discovery potential}
  \label{fig:sub2}
\end{subfigure}
\caption{The left plot shows the TS distribution of 0\%, 5\%, 50\% and 100\% of the Crab flux. The right plot shows the sensitivity and discovery potential curve.}

\label{fig:test2}
\end{figure}
Figure 4.a shows the TS distribution of different levels of flux. The redline in Figure 4.b shows the fraction of signal trials whose TS-value lies above the median of the background TS distribution. With 39\% of the Crab Nebula flux, 90\% of signal trials have a TS-value that is higher than the median of the background TS distribution. Similarly, the blue line shows the fraction of signal trials whose TS-value lies TS above the $3\sigma$ of the background TS distribution. With 81\% of the Crab Nebula flux, 50\% of the signal trials have TS-value that is higher than $3\sigma$ of the background TS distribution.

\section{Conclusions}
We present a python-based mximum likelihood software called i3mla. It is designed to be flexible and highly modularized to enable user-defined analyses. The compatibility with 3ML enable fast joint-fit analyses with other instruments and makes multi-messenger astronomy with IceCube coherent and easy. \\
We perform 3 different tests to validate i3mla. The signal and spectral index bias test shows i3mla can accurately fit the right flux and spectral index except for the known bias effect which will be fixed in the future. The flare identification test shows it can obtain a similar result to IceCube's previous TXS 0506+056 flare analysis. The HAWC-IceCube joint-fit test verifies the feasibility of a multi-messenger analysis.\\

\bibliographystyle{ICRC}
\bibliography{references}

\providecommand{\href}[2]{#2}\begingroup\raggedright\begin{thebibliography}{1}

\bibitem{Aartsen:2016nxy}
{\bfseries IceCube} Collaboration, M.~G. Aartsen {\em et~al.}
  \href{http://dx.doi.org/10.1088/1748-0221/12/03/P03012}{{\em JINST}
  {\bfseries 12} no.~03, (2017) P03012}.

\bibitem{Vianello:2015tuw}
G.~Vianello, R.~Lauer, P.~Younk, L.~Tibaldo, J.~M. Burgess, H.~Ayala~Solares,
  J.~P. Harding, C.~M. Hui, N.~Omodei, and H.~Zhou
  \href{http://dx.doi.org/10.22323/1.236.1042}{{\em PoS} {\bfseries ICRC2015}
  (2016) 1042}.

\bibitem{aartsen2018neutrino}
{\bfseries IceCube} Collaboration, M.~G. Aartsen {\em et~al.}
  \href{http://dx.doi.org/10.1126/science.aat2890}{{\em Science} {\bfseries
  361} no.~6398, (2018) 147--151}.

\bibitem{wilks1938large}
S.~S. Wilks {\em The Annals of Mathematical Statistics} {\bfseries 9} no.~1,
  (1938) 60--62.

\bibitem{abeysekara2017observation}
{\bfseries HAWC} Collaboration, A.~U. Abeysekara {\em et~al.}
  \href{http://dx.doi.org/10.3847/1538-4357/aa7555}{{\em Astrophys. J.}
  {\bfseries 843} no.~1, (2017) 39}.

\bibitem{abeysekara2019measurement}
{\bfseries HAWC} Collaboration, A.~U. Abeysekara {\em et~al.}
  \href{http://dx.doi.org/10.3847/1538-4357/ab2f7d}{{\em Astrophys. J.}
  {\bfseries 881} (2019) 134}.

\end{thebibliography}\endgroup
\clearpage
\section*{Full Author List: IceCube Collaboration}

% \noindent \textbf{Note comment afterwards:} Collaborations have the possibility to provide an authors list in xml format which will be used while generating the DOI entries making the full authors list searchable in databases like Inspire HEP. For instructions please go to icrc2021.desy.de/proceedings or contact us under icrc2021proc@desy.de.\\

% \scriptsize
% \noindent
% first.author$^1$, 
% second.author$^2$, 
% third.author$^3$ % .... more names
% and 
% last.author$^{n}$ \\

% \noindent
% $^1$first.affiliation.
% $^2$second.affiliation. % .... more affiliation
% $^{m}$last.affiliation.

\scriptsize
\noindent
R. Abbasi$^{17}$,
M. Ackermann$^{59}$,
J. Adams$^{18}$,
J. A. Aguilar$^{12}$,
M. Ahlers$^{22}$,
M. Ahrens$^{50}$,
C. Alispach$^{28}$,
A. A. Alves Jr.$^{31}$,
N. M. Amin$^{42}$,
R. An$^{14}$
% ...
\\

\noindent
% ...
$^{12}$ Universit{\'e} Libre de Bruxelles, Science Faculty CP230, B-1050 Brussels, Belgium \\
% ...
$^{14}$ Department of Physics and Laboratory for Particle Physics and Cosmology, Harvard University, Cambridge, MA 02138, USA \\
% ...
$^{17}$ Department of Physics, Loyola University Chicago, Chicago, IL 60660, USA \\
$^{18}$ Dept. of Physics and Astronomy, University of Canterbury, Private Bag 4800, Christchurch, New Zealand \\
% ...
$^{22}$ Niels Bohr Institute, University of Copenhagen, DK-2100 Copenhagen, Denmark \\
% ...
$^{28}$ D{\'e}partement de physique nucl{\'e}aire et corpusculaire, Universit{\'e} de Gen{\`e}ve, CH-1211 Gen{\`e}ve, Switzerland \\
% ...
$^{31}$ Karlsruhe Institute of Technology, Institute for Astroparticle Physics, D-76021 Karlsruhe, Germany  \\
% ...
$^{42}$ Bartol Research Institute and Dept. of Physics and Astronomy, University of Delaware, Newark, DE 19716, USA \\
% ...
$^{50}$ Oskar Klein Centre and Dept. of Physics, Stockholm University, SE-10691 Stockholm, Sweden \\
% ...
$^{59}$ DESY, D-15738 Zeuthen, Germany \\
% ...

\subsection*{Acknowledgements}

\noindent
USA {\textendash} U.S. National Science Foundation-Office of Polar Programs,
U.S. National Science Foundation-Physics Division,
U.S. National Science Foundation-EPSCoR,
Wisconsin Alumni Research Foundation,
Center for High Throughput Computing (CHTC) at the University of Wisconsin{\textendash}Madison,
Open Science Grid (OSG),
Extreme Science and Engineering Discovery Environment (XSEDE),
Frontera computing project at the Texas Advanced Computing Center,
U.S. Department of Energy-National Energy Research Scientific Computing Center,
Particle astrophysics research computing center at the University of Maryland,
Institute for Cyber-Enabled Research at Michigan State University,
and Astroparticle physics computational facility at Marquette University;
Belgium {\textendash} Funds for Scientific Research (FRS-FNRS and FWO),
FWO Odysseus and Big Science programmes,
and Belgian Federal Science Policy Office (Belspo);
Germany {\textendash} Bundesministerium f{\"u}r Bildung und Forschung (BMBF),
Deutsche Forschungsgemeinschaft (DFG),
Helmholtz Alliance for Astroparticle Physics (HAP),
Initiative and Networking Fund of the Helmholtz Association,
Deutsches Elektronen Synchrotron (DESY),
and High Performance Computing cluster of the RWTH Aachen;
Sweden {\textendash} Swedish Research Council,
Swedish Polar Research Secretariat,
Swedish National Infrastructure for Computing (SNIC),
and Knut and Alice Wallenberg Foundation;
Australia {\textendash} Australian Research Council;
Canada {\textendash} Natural Sciences and Engineering Research Council of Canada,
Calcul Qu{\'e}bec, Compute Ontario, Canada Foundation for Innovation, WestGrid, and Compute Canada;
Denmark {\textendash} Villum Fonden and Carlsberg Foundation;
New Zealand {\textendash} Marsden Fund;
Japan {\textendash} Japan Society for Promotion of Science (JSPS)
and Institute for Global Prominent Research (IGPR) of Chiba University;
Korea {\textendash} National Research Foundation of Korea (NRF);
Switzerland {\textendash} Swiss National Science Foundation (SNSF);
United Kingdom {\textendash} Department of Physics, University of Oxford.

\end{document}